\documentclass[creativecommons]{eptcs}
\providecommand{\event}{FORECAST 2016} 
\usepackage{breakurl}             

\title{On Formal Methods for Collective Adaptive\\System Engineering\\$ $\\
$
\left\{
\begin{array}{c}
\mbox{Scalable Approximated}\\
\mbox{Spatial}
\end{array}
\right\}
$
Analysis Techniques\\$ $\\$  $\\
{\large Extended Abstract}
}
\author{Diego Latella
\institute{CNR-ISTI\\ Pisa, Italy}
\institute{Istituto di Scienza e Tecnologie dell'Informazione ``A. Faedo''\\
Consiglio Nazionale delle Ricerche\thanks{Work partially funded by the EU projects QUANTICOL (nr. 600708) and ASCENS (nr. 257414).}\\
Pisa, Italy}
\email{Diego.Latella@cnr.it}
}
\def\titlerunning{Formal Methods for CAS}
\def\authorrunning{D. Latella}
\begin{document}
\maketitle

\begin{abstract}
In this extended abstract a view on the role of Formal Methods in System Engineering
is briefly presented. Then two examples of useful analysis techniques based 
on solid mathematical theories are discussed as well as the software tools which 
have been built for supporting such techniques. The first technique is
Scalable Approximated Population DTMC Model-checking. The second one is
Spatial Model-checking for Closure Spaces. Both techniques have been 
developed in the context of the EU funded project QUANTICOL.
\end{abstract}

\def\calB{{\cal B}}
\def\calC{{\cal C}}
\def\calM{{\cal M}}
\def\calN{{\cal N}}
\def\calP{{\cal P}}
\def\calS{{\cal S}}
\def\calX{{\cal X}}
\def\FlyFast#1{{\sf FlyFast#1}}

\section{Introduction}
When I was invited by the FORECAST organisers to give a broad overview of the current ``state of health''  of Formal Methods (FM), their usefulness and their actual use in (Collective Adaptive) System Engineering ((CA)SE), my first reaction was a gentle refusal, dictated by the  consideration that the audience at this workshop, and in general the SEFM Community,  is well aware of the current situation of FM.
So, we decided that, instead, I would have shared {\em my personal} view on the subject first,
and then I would have briefly described a couple of results  we have 
achieved in the context of the QUANTICOL Project (\url{http://www.quanticol.eu/}).

My professional experience induces me to think that the role of FMs in (CA)SE is still not fully appreciated in the Computer Science (CS) community, not even in the more intellectually sophisticated part thereof, namely in the CS research community\footnote{An extended version of this introduction can be found as two posts of mine in the blog of the ASCENS Project at \url{http://www.ascens-ist.eu/}.}, with a few exceptions like, of course, the FM Community itself and, maybe the Dependability one (or at least part of it).  So, I think that a few considerations on this subject might help the reader understanding why a great deal of 
effort in projects like QUANTICOL is devoted to the development of formal modelling and analysis languages and techniques.
My first and, I'd say major, observation is that this lack of appreciation for FMs seems to be a peculiarity of people's attitude towards SE, and is not at all experienced when considering other branches of engineering. Let me start by quoting C. Jones~\cite{Jon00}: ``All engineering disciplines make progress by employing mathematically based notations and methods.'' Note that the above consideration applies to all engineering disciplines, and it explicitly refers to notations and methods that are mathematically based. Is that true?

Let us consider civil engineering. In this case, graphical notations, among others, are widely used, and this goes on now for quite a long time.
Although the result of using such graphical notations can be quite evocative, and may bear some artistic value, it is worth noting that technical drawing is regulated by rather strict rules. Such rules are often formalised in international standards and are amazingly detailed (I still remember that, when I was an undergraduate student, we were taught, at a technical drawing course, that when you draw a certain kind of arrows, the width of the tip should be in a fixed relationship with its length---1/3, to be precise, if I remember well---in order for the arrow to be a ``correct'' one). Furthermore, we were taught specific techniques for constructing, or better, (almost) mechanically generating, sections of objects. Well, essentially, we were taught rigorous rules for drawing models of the artefacts we were designing, being them bridges or taps or electrical circuits. To a certain extent, these rules are a rigorous definition of the syntax to be used for creating our models. On the other hand, the sectioning/projecting techniques are based on formal rules of mechanical manipulation of the drawings, \ldots{} a, maybe rudimentary, form of model analysis based on formal semantics. 

Furthermore, in most cases, such drawings are decorated with numbers which could be used in mathematical formulae in close relation with (parts of) the drawings, or derivable from them,
in order to better analyse features of interest of the system one is designing. Take, for instance,
the blue-print of an electrical or electronic circuit design. 
Therein, we can identify a set of components composed according to precise rules (e.g.
in {\em series}, in {\em parallel}, in {\em series-parallel}, etc.), with precise {\em semantics}
associated to such composition rules, originating from a mathematical representation of physical phenomena (e.g. the resistance resulting by series composition of two or more resistors is the sum of the resistances of the component resistors, etc.). Once the link is established between our circuit, i.e. our design model, and its mathematical representation, one can proceed with sound mathematical manipulations which provide us with extremely valuable information. 
For instance, starting from the blue-print of an oscillator circuit we can derive 
an appropriate integral/differential equation using the values associated to the 
circuit components (for instance a resistor, a capacitor and an inductance in the case of
an electronic circuit). By studying the solution of the equation we can analyse important features of the circuit, like, for instance the {\em resonance} frequency which usually plays a fundamental
role in the behaviour of the system (which can even be dramatic in the case of mechanical
oscillating systems!). The description of these ``higher level'' features is typically the basis of
the ``technical specifications'' document, which is used for understanding whether the 
design satisfies important requirements. 

The process of relating a design model to the requirements specification can be seen as a process of abstraction and, of course, appropriate notions should be available also for checking that a design model satisfies a given requirement. But do we have to do all this by paper and pencil or only experimental testing? Fortunately not: we can play with software tools, even forgetting---while designing and only to a certain extent---that behind all these activities there is a huge and solid body of mathematics supporting them, including Set Theory, Continuous Mathematics, Metric Spaces, Differential Calculus and Function Analysis, Linear Algebra, Differential Equations, just to mention a few  \ldots

Summarising: Civil, Naval, Nuclear, Electrical, Electronic (\ldots) Engineers use notations for technical specifications (requirements specifications) as well as design specifications/models that
\begin{itemize}
\item    are strongly based on mathematics (and physics),
\item    are compositional (at least in a broad sense),
\item    are characterised by great and flexible descriptive power,
\item    allow for the formal manipulation of their objects,
\item    are heavily supported by computer (software) tools (e.g. for model analysis, including relating models to technical specs).
\end{itemize}
Society expects engineers to be aware of the underlying theories although not necessarily able to completely master them. I repeat it: a good, or even average, civil, electrical, etc. engineer, is socially not allowed to claim he can do without formal methods. Actually, the situation
is even better for traditional engineering. In fact the basics of the methods and underlying maths are even taught to high school students (at least in technical professional schools)! 
And now: what about engineers designing complex, critical, computer/software systems? Although in the past there have been people, including respectable scientists, who have been arguing that SE is mainly a matter of art, or, at most of craftwork, there is now some 
consensus---at least in the SE community itself and thanks to the 30 years long tradition of 
research, dissemination, and education carried out within and by the FM Community---that it is a matter of engineering and so, as in any other branch of engineering, besides ingenuity and inspiration, there is need of {\em systematic application of sound methodologies and techniques, with a solid mathematical basis}.   In addition, 
the mathematical basis as well as the methodologies must be part of standard 
education curricula of future Computer System Engineers (and Practitioners, I would add!) and
this is even more important when we focus on CAS.
These systems consist of a large number of spatially distributed heterogeneous entities with decentralised control and varying degrees of complex autonomous behaviour. 
Often humans act both as agents within the system and as end-users, outside the system. 
It is worth noting that as end-users, they are not necessarily aware of the sophisticated underlying technology and the complexity of CAS. On the other hand, given the fact 
that such systems are meant to support critical  socio-technical infrastructures, such as
transportation and energy distribution ones, it is necessary that 
 thorough a priori analysis of their design is carried out to investigate all aspects of their behaviour, including quantitative and emergent aspects, before they are put into operation.

In the sequel I'll sketch two examples of formal analysis techniques we developed
in the context of the QUANTICOL Project, using non-trivial mathematical notions and results
which address two important issues
in the field of CAS, namely {\em scalability} of analysis~\cite{LLM14,LLM15a,LLM15b} and reasoning about {\em space}~\cite{Ci+14a}. For further information on CAS and related work see e.g. the web site of the QUANTICOL project at \url{http://www.quanticol.eu}, and that of the FOCAS Coordination Action at \url{http://www.focas.eu}.

The work briefly described  below has been carried out jointly mainly with Vincenzo Ciancia\footnote{\label{ISTI} CNR-ISTI.}, Michele Loreti\footnote{ Univ. di Firenze and IMT-Lucca.}, and Mieke Massink$^{\scriptsize\ref{ISTI}}$.

\section{Scalable Approximated Population DTMC Model-checking}
Model-checking is a powerful approach to the automatic verification of complex (computer) systems. It consists of an efficient procedure that, given an abstract model $\calM$ of the system, decides whether $\calM$ satisfies a (typically temporal, but also probabilistic) logic formula.
Unfortunately, model-checking procedures suffer of the potential combinatorial explosion of the state space so that their scalability is a serious concern. The applicability
of model-checking techniques to CAS is thus a particularly serious problem given that
one of the characterising features of such systems is their large or even huge size. 

In~\cite{LLM14,LLM15a} we proposed a novel model-checking procedure for
probabilistic Discrete Time Markov Chains (DTMC) based population models which 
is insensitive to the size of the population the system consists of and is thus scalable. The procedure is based on an original combination of local, on-the-fly probabilistic model-checking techniques and mean-field approximation in discrete time~\cite{BMM07}. The procedure can be used to verify bounded PCTL~\cite{Hansson1994} properties of selected individuals in the context of systems consisting of a large number of similar but independent interacting objects. 

The asymptotic correctness of the model-checking procedure has been proven and a prototype implementation of the model-checker, \FlyFast{,} has been applied to several benchmark examples.  To the best of our knowledge, this is the first implementation of an {\em on-the-fly mean field} model-checker for {\em discrete time, probabilistic, time-synchronous} models.

The approach we followed is similar to the one proposed in~\cite{BMM07} for {\em fast simulation} (thence the ``Fast'' part of the name of our tool). A model for $N$ interacting objects is considered
where the evolution of each object is given by a finite state DTMC, the transition matrix of which
may depend on the distribution of states of all the objects in the system\footnote{ Strictly speaking, thus, this DTMC is time-inhomogeneous.}. Each object can be in one of its local states at any point in time and all objects proceed in discrete time and in a clock-synchronous fashion.
The distribution of states of all objects in the system at a given step of the computation is represented by the so-called {\em occupancy measure} vector. The size $S$ of this vector is equal to the total number of the states of the objects in the system; each element stores the {\em fraction}---over the total number of objects---of the objects in the system which are currently in the corresponding state. The overall system behaviour is a $S^N \times S^N$ DTMC, hence the state-space explosion problem. Obviously, also the evolution in time of the occupancy measure vector
is a DTMC. As shown in~\cite{BMM07}, when the number of objects is sufficiently large, the latter can be approximated by the {\em deterministic} solution of a {\em difference equation} which is called the {\em mean field}.  This convergence result has been exploited in~\cite{BMM07} to obtain a `fast' way to stochastically simulate the evolution of a selected, limited number of specific objects in the context of the overall behaviour of the population. 

We showed that the deterministic iterative procedure of~\cite{BMM07} for approximating the occupancy measure combines well with an {\em on-the-fly} (thence the ``Fly'' part of the name of our tool) probabilistic model-checking procedure for the verification of bounded PCTL formulae addressing selected objects of interest\footnote{ Note that the transition probabilities of these selected objects at time $t$ may depend on the occupancy measure of the system at $t$ and therefore also the truth-values of the formulae may vary over time.}.  An on-the-fly recursive approach also provides a natural way to address nested path formulae and time-varying truth values of such formulae.  The considered PCTL formulae can be extended with properties that address the overall status of the system. The use of the mean-field semantics instead of
the standard probabilistic semantics as the underlying semantic model for the  on-the-fly
procedure produces a dramatic decrease in the number of possible next states to consider 
in each expansion step of the algorithm. For instance, in the typical case in which one is interested in analysing the behaviour of one object (with $S$ states) in the context of a system of
$N$ objects, for large $N$, the number of possible next states decreases  from $S^N \times S^N$ to just $S$.  Examples of application of the scalable approximated population DTMC model-checking technique sketched above can be found in~\cite{LLM13b,LLM14,LLM15a,LLM15b,CLM16a}; these include classical CAS examples like
computer epidemic and bike-sharing systems, the Predator-prey model 
of Lotka-Volterra 
as well as approximation of
fluid, population CTMC model-checking based on discrete time---as opposed 
to the standard, continuous time approach mentioned below.

The work outlined in this section is based on {\em discrete time}, 
clock-synchronous, probabilistic models of computation. In~\cite{BoH12b,BoH15} a similar problem is tackled, but in the
{\em continuous time}, interleaving and stochastic context. In particular, a fluid-flow
semantics is used for the system population, the overall behaviour of which is approximated by
the solution of a set of {\em differential equations}. In this context, the behaviour
of the selected individual(s) is modelled by a (time-inhomogeneous) Continuous Time 
Markov Chain and the properties of interest are expressed in CSL~\cite{Az+00,Ba+03}.
Preliminary ideas on the exploitation of mean-field convergence in continuous time for model- checking mean-field models, and in particular for an extension of the logic CSL, were informally sketched in a presentation at QAPL 2012~\cite{Ko+12}, but no model-checking algorithms were presented. Follow-up work on the above mentioned approach can be found in~\cite{Ko+13} which relies on the earlier results presented in~\cite{BoH12b}.
We are not aware of other contributions in the literature addressing 
scalability of model-checking using mean-field/fluid-flow semantics.

\section{Automatic Reasoning About Space}
In the previous section it has been underlined that one of the characterising feature of CAS 
is that they are large systems. Another feature of such systems is that their global behaviour critically depends on interactions which are often local in nature. The aspect of locality immediately poses issues of spatial distribution of objects. There are cases in which the analysis of features
of such systems simply {\em cannot} abstract from space. For example, consider a bike (or car) sharing system having several parking stations, and featuring twice as many parking slots as there are vehicles in the system. Ignoring the spatial dimension, on average, the probability to find completely full or empty parking stations at an arbitrary station is very low; however, this kind of analysis may be misleading, as in practice some stations are placed in places which 
are much more popular than others. 
Consequently, the probability of finding a station full or empty heavily depends on the location of  the station.
In such situations, it is important to be able to formally reason about spatial properties
of CAS and, possibly, to do this with the help of automatic software tools. 

In the tradition of mathematical logic, there is a great deal of literature dealing
with the issue of formalisation of space and of spatial reasoning. In particular, 
\emph{spatial} logics take the approach of a spatial interpretation of modal logics.
Dating back to early logicians such as Tarski, it has been shown that 
modalities may be interpreted using the concept of \emph{neighbourhood} in a topological space
and it has been discovered that classical axiomatisations of modal logic like S4 
work perfectly well also for the topological interpretation~\cite{Ai+07}.

The field of spatial logics is well developed in terms of descriptive languages and computability/complexity aspects.  However, most of the proposals in the literature are focussed on continuous
models, leaving discrete ones rather unexplored. In addition, and related to this, 
verification issues, and in particular model-checking, are not addressed, to the best of our
knowledge.

In~\cite{Ci+14a} we take the approach of Galton~\cite{Gal03} and use \emph{closure spaces}
as the main domain for space. Closure spaces generalise the notion of topological 
spaces in the way briefly described in the sequel. One of the existing definitions for topological space is that such a space is a pair $(X,\calC)$ where $X$ is a set and the
{\em closure operator} $\calC: 2^X \rightarrow 2^X$ is a function which has to satisfy the following four axioms for all $A,B \subseteq X$:
\begin{enumerate}
\item $\calC(\emptyset) =\emptyset$;
\item $A \subseteq \calC(A)$;
\item $\calC(A \cup B) =\calC(A) \cup \calC(B)$;
\item\label{IDEM} $\calC(\calC(A)) =\calC(A)$.
\end{enumerate}

In the case of Euclidean spaces, $\calC(X)$ is exactly the
standard closure of $X$; for example we have $\calC((0,1)) = [0,1]$.

If we remove the idempotence requirement (axiom \ref{IDEM}), we get a closure space.
In addition, it can be seen that any relation $R\subseteq X \times X$ immediately
induces a closure operator $\calC_R$.
So, in this approach, graphs and topological spaces are treated in a uniform framework. 

We defined a spatial logic, SLCS (Spatial Logic of Closure Spaces), the formulae of which are interpreted over points $x$ in a closure space $\calX.$
Besides standard conjunction and negation operators we introduced a `near' operator $\calN$
and a `surrounded' operator $\calS$
such that:\\
\begin{itemize}
\item[] $x \models_{\calX} \calN \phi$ iff $x \in \calC(\{y| y \models_{\calX} \phi\})$, and 
\item[] $x \models_{\calX} \phi \calS \psi$ iff 
there exists set $A\subseteq X$ s.t.  $x \in A$,
$A \subseteq \{y| y \models_{\calX} \phi\}$ and 
$(\calC(A) \setminus A) \subseteq \{z| z \models_{\calX} \psi\}$.\\
\end{itemize}

$\calS$ is a spatial interpretation of the fundamental \emph{until} operator from temporal logic. Intuitively, $\phi \calS \psi$ describes a situation in which it is not possible to `escape' an area of points satisfying $\phi$, without passing through at least one point that satisfies $\psi$. 
To formalise this intuition, we provided also a characterising theorem that relates infinite paths in a closure space and surrounded formulae. 

This small set of operators is surprisingly
expressive. For example, a number of interesting derived operators can
be defined, including the well-known spatial `somewhere' and
`everywhere' operators, and various forms of reachability.
We finally introduced an efficient model-checking procedure and produced an implementation of a spatial model-checker called {\tt topochecker}; the tool is able to interpret spatial logics on generic graphs 
including digital images, providing graphical understanding of the meaning of formulae, and an immediate form of counterexample visualisation. 
In \cite{Ci+15b} the logic and the 
model-checking procedure have been extended with classical temporal logic operators
so that a spatio-temporal logic and model-checking tool is now available for reasoning about dynamically changing spaces give rise.
For example, one can characterise those points which {\em will eventually} be
surrounded by points which {\em may eventually} be close to points satisfying $\phi$.
Examples of application of SLCS model-checking, including its spatio-temporal extension
can be found in~\cite{Ci+14a,Ci+14c,Ci+16,Ci+16a,Ci+16b}; these include spatio-temporal model-checking of (digital images of) vehicular movement in public transport systems including bus clumping avoidance,
reachability of exits in a maze,  emergency building evacuation and other rescuing operations,
and the study the emergence of patterns in bio-chemical systems.

We are not aware of other published work on spatial or spatio-temporal model-checking with 
the exception of~\cite{Gr+08,Gr+09,Ha+15,Ne+15}. 

In~\cite{Gr+08,Gr+09} a variant of spatial logic is proposed in which spatial properties are expressed using ideas from image processing, namely quad trees. This variant is equipped with practical model checking algorithms and with machine learning procedures and allows one to capture very complex spatial structures. However, this comes at the price of a complex formulation of spatial properties, which need to be learned from some template image. The combination of this spatial logic with linear time signal temporal logic, defined with respect to continuous-valued signals, has recently led to the spatio-temporal logic SpaTeL~\cite{Ha+15}.

In~\cite{Ne+15}  the Spatial Signal Temporal Logic is presented, which is based on the $\calS$ 
operator described above, extended with a notion of distance. A monitoring algorithm 
is presented as well, which includes  spatial model-checking functionality as well.

\section{Conclusions}
In this extended abstract, after some general personal considerations on the role of
Formal Methods in System Engineering, two examples
of useful techniques for the analysis of important features of CAS
have been briefly presented which, although both giving rise to tools belonging to the category 
of ``push-button'' tools, are based on deep mathematical theories; specifically
a convergence theory for stochastic processes and the theory of topological and closure spaces.
Our future plan is to integrate the results mentioned above. 

For instance, it has been shown that an appropriate interaction paradigm for CAS 
is one which allows system components to communicate using multi-cast send/receive
primitives where the partners of interest are identified by means of {\em predicates}
on dynamic features of the components, typically represented by the current values of specific
component {\em attributes}~\cite{De+15,LoH16}. To that purpose, each component
is typically constituted by a process behaviour and by a {\em store} where attribute values
are kept. So, in~\cite{CLM16a}
we provide a front-end language for \FlyFast{} allowing for the use of
object attribute-/predicate-based communication, where attributes can take
points in closure spaces as values, among others,  and simple predicates can be used
for controlling object communication. 
Another challenging line of research would be to extend the predicate based interaction paradigm with the possibility of using SLCS (and its extension's) formulae as predicates
in component interaction. Similarly challenging would be to provide components 
with reasoning capabilities using their store. For example, in~\cite{CLM15}
we show an example where closure spaces are used not only for modelling
the space where components live, but also for supporting reasoning capabilities of
individual components about (their location in) space during their own execution. 
Such reasoning capabilities are quite primitive in the example mentioned above; it would be  interesting to investigate more sophisticated reasoning techniques, including spatial model-checking, in the general component and system modelling/verification techniques 
and tool support framework.

A closely related research line which we have started recently is the combination
of {\em statistical} model-checking~\cite{LaL14,SeV13} with spatio-temporal model-checking.
An example of application of the combined technique to the analysis of
a bike-sharing system of the size of the London one is presented in~\cite{Ci+16x}.
An application in the area of medical imaging is reported in a paper elsewhere in these
proceedings.

Further work is also required  at the level of the spatial logic itself. We have recently started
working on an additional operator $\calP$: intuitively the formula $\phi \calP \psi$ describes a situation in which the points satisfying $\psi$ can be reached by paths rooted in points satisfying $\phi$ and, for the rest, composed only of points satisfying $\psi$; we furthermore are 
extending the logic with operators for collective properties, namely properties which are satisfied by connected sets of points, rather than points in isolation; in addition, we are
investigating a alternative
definitions of the satisfaction relation, based on the notion of spatial {\em paths}, which we
hope pave the way for the application of the logic in reasoning about continuous space;
the interested reader is referred to~\cite{Ci+16a}.

%

\end{document}